\documentstyle[12pt,twoside]{article}
\begin{document}
\begin{center}

{\Large\bf QUANTUM COSMOLOGY\\[5pt]
IN SCALAR-TENSOR THEORIES\\[5pt]
WITH NON MINIMAL COUPLING\\[5pt]}
\medskip

{\bf J\'ulio 
C. Fabris\footnote{e-mail: fabris@cce.ufes.br}}
\medskip

Departamento de F\'{\i}sica, Universidade Federal do Esp\'{\i}rito Santo, 
29060-900, Vit\'oria, Esp\'{\i}rito Santo, Brazil
\medskip
\\
{\bf Nelson Pinto-Neto\footnote{e-mail: nen@lca1.drp.cbpf.br}}
and {\bf A. F. Velasco}
\medskip

Lafex, Centro Brasileiro de Pesquisas F\'{\i}sicas, Rua Xavier Sigaud 150,
22290-180, Rio de Janeiro, Brazil

\end{center}

\begin{abstract}
Quantization in the minisuperspace of non minimal scalar-tensor theories
leads to a partial differential equation which is non separable.  Through a
conformal transformation we can recast the Wheeler-DeWitt equation in an
integrable form, which corresponds to the minimal coupling case, whose
general solution is known. Performing the inverse conformal transformation
in the solution so found, we can construct the corresponding one in the
original frame. This procedure can also be employed with the bohmian
trajectories.  In this way, we can study the classical limit of some
solutions of this quantum model. While the classical limit of these
solutions occurs for small scale factors in the Einstein's frame, it
happens for small values of the scalar field non minimally coupled to
gravity in the Jordan's frame, which includes large scale factors.
\vspace{0.7cm}
PACS number(s): 04.20.Cv., 98.80.Hw
\end{abstract}

\section{Introduction}

The theories of unification of interactions, like Kaluza-Klein, supergravity
or superstring theories, predict non minimal couplings between the geometry
of spacetime and a scalar field. The covariant quantization of these theories
in term of perturbative methods is incomplete and it is interesting to pursue
non-perturbative methods or simplifications in order to obtain some qualitative
features of the quantization of these theories. The most dramatic simplification
is the minisuperspace quantization, where all but a finite number of degrees of
freedom are frozen yielding, in general, a solvable quantum problem.
The interest on these models appears because, in spite of all this simplification,
there are some issues which still persist, like the issue of time, how to
interpret the wave function of the Universe, the elimination of singularities
by quantum effects, and the conditions for the classical limit. 

In the present paper, we will study these problems in the context of
minisuperspace quantization of scalar-tensor theories with non minimal
coupling. These theories are derived from the effective action of
string theories \cite{barrow}, and from Kaluza-Klein theories
compactified to four dimensions \cite{multi,julio1}. In both cases, the
resulting model has a non minimal coupling between gravity and one or
more scalar fields. A conformal transformation permits to
write these effective models in the Einstein's frame, where the scalar
field is minimally coupled to the gravity sector. There are claims that
these minimally coupled formulations correspond to the physical frame in
such a way that the conformal transformation does not constitute a
mathematical tool only \cite{gunzig,levin}; to our point of view this
is not a closed subject, and the choice of a physical frame depends on
the theoretical structure one has in mind, for example, if we intend to
consider a constant or a variable gravitational coupling. In some
cases, stability considerations may select a prefered frame
\cite{sokolowsky}.

In a recent paper \cite{nelson1}, the minisuperspace quantization of scalar-tensor
theories with minimal coupling was studied. The presence of singularities
at the quantum level, the classical limit, the issue of time, and problems
of interpretation were discussed. It was shown that the causal interpretation
of quantum mechanics \cite{boh}, one of the possible interpretations which can be used in quantum cosmology (for reviews on interpretations and classical 
limit in
quantum cosmology, see Ref. \cite{nelson}),
where the issue of time is absent at the minisuperspace level
\cite{nelson2}, 
yields a classical limit which is in agreement with semiclassical
considerations. The general solution of the Wheeler-DeWitt equation was found.
For some particular exact solutions, 
it was shown that, contrary to common belief, the classical
limit occurs for small scale factors, and hence the classical singularities
are not avoided by quantum effects, which become important only for larger
values of the scale factor. For the case of  
scalar-tensor theories with non minimal coupling, the corresponding
Wheeler-DeWitt equation is non separable and more difficult to solve.
However, we were able to show that not only the classical solutions but also
the quantum theory of this model can be mapped to the minimal coupling
case by a conformal transformation. The solutions of the Wheeler-DeWitt equation
found in the Einstein's frame can be mapped to the solutions of the
more complicated Wheeler-DeWitt equation in the Jordan's frame,
and also the bohmian trajectories which describe the quantum evolution 
of the solutions. 

At the classical level, the effective models considered here are labeled by
a coupling constant $\omega$, which in principle may take values in the
interval $- \frac{3}{2} < \omega < \infty$, as in the Brans-Dicke
theory. The models that will be studied here differ from
the usual Brans-Dicke theory by the presence of a second scalar field
minimaly coupled to gravity but with a non trivial coupling
with the usual scalar field of the Brans-Dicke theory. This  extra
scalar field is
usually present in the effective action quoted above. For string effective action $\omega = - 1$, while for
compactified Kaluza-Klein theories, $\omega = - \frac{d - 1}{d}$, where
$n=4+d$ is the dimension of space-time. In this paper, we will consider
only spatial sections of the Friedmann's model with positive
curvature.  For $\omega < 0$, the classical solutions represent 
universes with bounce (but not necessarily singularity-free), while
for $\omega > 0$, the cosmological scenario is qualitatively like the
traditional Friedmann's cosmological scenarios. For $\omega = 0$, we
have a remarkable oscillating Universe, where the scale factor never
goes to zero, constituting a complete singularity-free model.  
\par 
At the quantum level, 
the bohmian trajectories of the quantum solutions studied in this paper
show the same behavior as the classical
solutions whenever the scalar field non minimally coupled to gravity is small
(with the exception of the very particular case $\omega =0$), which coincides with small 
values of the scale factor $a$ when $\omega > 0$, and large values of $a$ when 
$-\frac{3}{2} > \omega  > 0$. Hence, contrary to the minimal
coupling case,
the bohmian trajectories may coincide with the classical solutions for
large values of the scale factor\footnote{We would like to emphasize that these comparisons and
results were 
obtained for some particular exact solutions of the Wheeler-DeWitt equation
in both frames, which can be written in term of elementary functions.
For more complicate solutions, the classical limit may present other 
features. However, these examples are sufficient to show that the quantum
theory in Einstein's and Jordan's frames have quite different properties.}.  
\par 
In the next section we exhibit the classical
theory, with their corresponding classical solutions. In section $3$ we
show the conformal mapping of the Wheeler-DeWitt equation in the
mini-superspace from Jordan's to Einstein's frames, from where we
determine their solutions in the non minimal coupling case.  In section
$4$, the conformal equivalence of the bohmian trajectories is also
determined, and their classical limits
are compared and discussed. The conclusions are presented in section $5$.

\section{Scalar-tensor theories}

In absence of ordinary matter, the most general Lagrangian density we can
write
for scalar-tensor theories is,
\begin{equation}
\label{l1}
{\it L} = \sqrt{-g}\biggr(f(\phi)R -
\omega(\phi)\frac{\phi_{;\rho}\phi^{;\rho}}{\phi} +
V(\phi)\biggl) \quad ,
\end{equation}
where $f(\phi)$ and $\omega(\phi)$ are arbitrary functions of the scalar
field,
while $V(\phi)$ is a potential term.
In general the potential term may be added by hand, although in some
cases the
choice can be dictated by microphysical considerations.
The Brans-Dicke theory \cite{brans}
is represented by the special case where $f(\phi) = \phi$ and
$\omega(\phi) = \mbox{constant}$; generally, in Brans-Dicke theory, the
potential
term is put equal to zero.
\par
Scalar-tensor theories appear in the low energy limit
of string theory, and in the reduction of Kaluza-Klein theories to four
dimensions.
The form of these effective Lagrangian depends on the way the
compactification is
made, and on the original multidimensional theories (as in supergravity
theories).
In cosmology, generally, we are interested only in the bosonic sector
without
gauge fields in four dimensions. A large class of these effective theories
is represented by a Lagrangian of the type
\begin{equation}
\label{l22}
{\it L} = \sqrt{-g}\biggr(\phi R -
\omega\frac{\phi_{;\rho}\phi^{;\rho}}{\phi}
- \frac{\chi_{;\rho}\chi^{;\rho}}{\phi}\biggl) \quad .
\end{equation}
This Lagrangian is the same as the Brans-Dicke one, but it has an
additional scalar field,
which couples non trivially with the Brans-Dicke field $\phi$.
Such effective Lagrangian appears in string cosmology with axion field
($\omega = - 1$),
and in multidimensional theories where gravity is coupled to external (non
geometric)
gauge fields in the higher dimensional space-time $n = 4 + d$ ($\omega = -
(d - 1)/d$).
\par
Employing the variational principle, we find the field equations
corresponding to this
Lagrangian:
\begin{eqnarray}
\label{fe1}
R_{\mu\nu} - \frac{1}{2}g_{\mu\nu}R &=&
\frac{\omega}{\phi^2}\biggr(\phi_{;\mu}\phi_{;\nu}
- \frac{1}{2}g_{\mu\nu}\phi_{;\rho}\phi^{;\rho}\biggl) +
\frac{1}{\phi}\biggr(\phi_{;\mu;\nu} - g_{\mu\nu}\Box\phi\biggl) \nonumber
+\\
&+& \frac{1}{\phi^2}\biggr(\chi_{;\mu}\chi_{;\nu} -
\frac{1}{2}g_{\mu\nu}\chi_{;\rho}\chi^{;\rho}\biggl) \quad ;\\
\nonumber\\
\label{fe2}
\Box\phi + \frac{2}{3 + 2\omega}\frac{\chi_{;\rho}\chi^{;\rho}}{\phi} &=&
0 \quad ; \\
\nonumber \\
\label{fe3}
\Box\chi - \frac{\phi_{;\rho}}{\phi}\chi^{;\rho} &=& 0 \quad .
\end{eqnarray}
\par
We insert in the field equations (\ref{fe1},\ref{fe2},\ref{fe3}) the
Friedmann-Robertson-Walker metric
\begin{equation}
\label{metric}
ds^2 = N^2dt^2 - a(t)^2\biggr(\frac{dr^2}{1 - kr^2} + r^2(d\theta^2 +
\sin ^2\theta d\phi^2)\biggl)
\end{equation}
where $a$ is the scale factor of the Universe, and $k = 0,1,-1$ represents
the constant curvature of the spatial section. The factor $N$ is the
lapse function, which we will set
equal to one in this section. The resulting equations of motion are:
\begin{eqnarray}
\label{em1}
3\biggr(\frac{\dot a}{a}\biggl)^2 + 3\frac{k}{a^2} &=&
\frac{\omega}{2}\biggr(\frac{\dot\phi}{\phi}\biggl)^2
- 3\frac{\dot a}{a}\frac{\dot\phi}{\phi} +
\frac{1}{2}\biggr(\frac{\dot\chi}{\phi}\biggl)^2 \quad ;\\
\nonumber\\
\label{em2}
\ddot\phi + 3\frac{\dot a}{a}\dot\phi + \frac{2}{3 +
2\omega}\frac{{\dot\chi}^2}{\phi} &=& 0 \quad ; \\
\nonumber \\
\label{em3}
\ddot\chi + 3\frac{\dot a}{a}\dot\chi - \frac{\dot\phi}{\phi}{\dot\chi} &=& 0
\quad .
\end{eqnarray}
The dot means derivative with respect to the cosmic time $t$. To solve
the equations (\ref{em1},\ref{em2},\ref{em3}) it is easier to
reparametrize the
time coordinate as $dt = a^3d\theta$.
In terms of $\theta$, we obtain the equations of motion,
\begin{eqnarray}
\label{em4}
3\biggr(\frac{a'}{a}\biggl)^2 + 3ka^4 &=&
\frac{\omega}{2}\biggr(\frac{\phi'}{\phi}\biggl)^2
- 3\frac{a'}{a}\frac{\phi'}{\phi} +
\frac{1}{2}\biggr(\frac{\chi'}{\phi}\biggl)^2 \quad ;\\
\nonumber\\
\label{em5}
\phi'' + \frac{2}{3 + 2\omega}\frac{{\chi'}^2}{\phi} &=& 0 \quad ; \\
\nonumber \\
\label{em6}
\chi'' - \frac{\phi'}{\phi}\chi' &=& 0 \quad .
\end{eqnarray}
The prime means derivative with respect to $\theta$.
\par
The solution of equation (\ref{em6}) is direct and reads,
\begin{equation}
\label{cs1}
\chi' = C\phi \quad ,
\end{equation}
where $C$ is an integration constant. Inserting (\ref{cs1}) into
(\ref{em5}) we obtain an harmonic oscillator
equation for $\phi$ with the solution,
\begin{equation}
\label{cs2}
\phi = D\sin(\lambda\theta) \quad ,
\end{equation}
$D$ being another integration constant and $\lambda = \sqrt{\frac{2}{3 +
2\omega}}C$. Now, we turn to equation (\ref{em4}). We insert in it the
solutions (\ref{cs1},\ref{cs2}). Redefining the scale factor
$a = \phi^{-1/2}b$, we can integrate the resulting equation for $b$. This
redefinition is the same that
transforms the original Lagrangian written in the Jordan's frame to the
equivalent Lagrangian written in the Einstein's frame, whose cosmological
solutions for
a closed spatial section has been solved previously \cite{nelson1}. The
final solution for the scale factor is:
\begin{equation}
\label{cs3}
a=\sqrt{\frac{2Cr\tan^{\alpha}(\frac{\lambda\theta}{2})}{\sqrt{6}\sin(\lambda\theta)[1 + r^2\tan^{2\alpha}(\frac{\lambda\theta}{2})]}} \quad ,
\end{equation}
where $\alpha \equiv \sqrt{1+\frac{2}{3}\omega}$, and $r$ is an integration constant.
When $\omega > 0$, (\ref{cs3}) represents an Universe that has an
expanding phase, coming from a singularity,
reaching a maximum value for $a$, and then collapsing again to a
singularity. For $\omega < 0$, the scenario
is completely different: the Universe has initially a contracting phase coming from
$a\rightarrow\infty$, reaches a minimum, and then
enter in an expansion phase until $a\rightarrow\infty$. This bouncing Universe 
is free of
singularities when
$- \frac{3}{2} < \omega < - \frac{4}{3}$, with deflationary and inflationary
periods when $a\rightarrow\infty$ (in the case where $w=4/3$ this inflation is
exponential), otherwise there are 
singularities when $a\rightarrow\infty$. When $\omega > 0$ all energy
conditions are satisfied, while in the
second case, the strong energy condition is always violated and the weak
energy conditions
can be violated in some specific regions. In the limits $\theta \rightarrow
0$ and $\theta \rightarrow \pi /\lambda$, the solutions 
take the form,
\begin{equation}
\label{roni}
a \propto t^{\frac{\alpha -1}{3\alpha -1}}, \quad \phi \propto t^\frac{2}{3\alpha
- 1}, \quad \chi \propto {\rm const.},
\end{equation}
with $t \propto \theta ^{(3\alpha - 1)/2}$ when $\theta \rightarrow 0$ or
$t \propto (\pi - \lambda\theta) ^{(3\alpha - 1)/2}$ when $\theta \rightarrow 
\pi /\lambda$
For $\omega > 0$ there is no
inflationary phase.
In the case  $- \frac{3}{2} < \omega < - \frac{4}{3}$ ($0 < \alpha < \frac{1}{3}$),
$\theta \rightarrow 0$ implies $t \rightarrow - \infty$, while
in all other cases $\theta \rightarrow 0$
implies $t \rightarrow 0$.
\par
The solutions with big-crunch following the big-bang ($\omega > 0$), and
those with bounce ($\omega < 0$),
are separated by the particular case where $\omega = 0$, for which the
scale factor oscillates between
a maximum and a minimum value, taking the form,
\begin{equation}
\label{cs}
a =\sqrt{\frac{Cr}{\sqrt{6}(\cos^2(\frac{\lambda\theta}{2}) + 
r^2\sin^2(\frac{\lambda\theta}{2}))}}
\quad ,
\end{equation}
This is a non singular solution. In the particular case where
$r = 1$, the scale factor
is constant. It is an static Universe,
even with the scalar fields evolving in time.
\par
It is worth to remember the corresponding solutions for the minimal
coupling case, which can be obtained
by performing the transformation $a = \phi^{-1/2}b$, yielding
\begin{eqnarray}
b &=& \sqrt{\lambda _1 \mbox{sech}(2\lambda _1\epsilon)}
\quad , \\
\phi &=& D\mbox{sech}\biggr(\frac{2\lambda _1}{\alpha}\epsilon\biggl) \quad ,
\\
\chi &=& \sqrt{\frac{3}{2}}\alpha D\mbox{tanh}\biggr(\frac{2\lambda _1}{\alpha}
\epsilon\biggl) + \chi _0  \quad ,
\end{eqnarray}
where $d\theta = \phi d\epsilon$, $\lambda _1 \equiv CD/\sqrt{6}$, and $\chi _0$ is an integration
constant. Near the singularities, the above solutions in the
Einstein's frame
behave as $b \propto \tau^{1/3}$, $\phi \propto \tau^{2/3\alpha}$
and $\chi \propto {\rm const.}$, where $\tau$ is the proper time in that
frame, d$\tau=\phi ^{1/2}$d$t$.

\section{Quantum solutions in the minisuperspace}

We now insert the metric (\ref{metric}) in the Lagrangian (\ref{l22}).
After integration by parts, we obtain the
expression,
\begin{equation}
\label{l2}
{\it L} = \frac{1}{N}\biggr(12a\dot a^2\phi + 12a^2\dot
a\dot\phi
- 2\omega\frac{\dot\phi^2}{\phi}a^3 - 2\frac{\dot\chi^2}{\phi}a^3\biggl)
- 12kNa\phi
\quad ,
\end{equation}
The conjugate momenta are,
\begin{eqnarray}
\pi_a &=& \frac{\partial L}{\partial\dot a} =
12\frac{a^2\phi}{N}\biggr(2\frac{\dot a}{a} +
\frac{\dot\phi}{\phi}\biggl) \quad , \\
\pi_\phi &=& \frac{\partial L}{\partial\dot\phi} =
4\frac{a^3}{N}\biggr(3\frac{\dot a}{a} -
\omega\frac{\dot\phi}{\phi}\biggl) \quad , \\
\pi_\chi &=& \frac{\partial L}{\partial\dot\chi} =
-4\frac{a^3}{N\phi}\dot\chi \quad .
\end{eqnarray}
The lapse function $N$ is a Lagrangian multiplier whose variation leads to a 
constraint equation.
The Hamiltonian can be obtained in the canonical way, and it reads,
\begin{equation}
H = N{\cal H}\quad ,
\end{equation}
with
\begin{equation}
{\cal H} = \frac{a^3}{\phi}\biggr(\frac{\omega}{24(3+2\omega)}\frac{\pi_a^2}{a^4}
+ \frac{\phi}{4(3+2\omega)}\frac{\pi_a\pi_\phi}{a^5}
- \frac{1}{4(3+2\omega)}\frac{\phi^2}{a^6}\pi_\phi^2 -
\frac{\phi^2}{8a^6}\pi_\chi^2 + 12\frac{\phi^2}{a^2}\biggl) \quad .
\end{equation}
Since the function $N$ is a Lagrange multiplier, we have the
constraint,
${\cal H} \approx 0 $. Applying the Dirac quantization procedure,
the quantum states must be annihilated by the operator version of ${\cal H}$,
yielding,
\begin{equation}
{\hat{\cal H}}\Psi = 0
\end{equation}
where $\Psi$ is the wavefunction of the Universe.
Inserting the explicit operator version for ${\hat{\cal H}}$ in the above equation by substituting $\pi_a = -
i\partial_a$, $\pi_\phi = - i\partial_\phi$,
$\pi_\chi = - i\partial_\chi$,
we obtain the differential equation for $\Psi = \Psi(a,\phi,\chi)$:
\begin{equation}
\label{wdw1}
\biggr\{\frac{\omega}{6(2\omega + 3)}a^2\biggr(\partial^2_a +
\frac{p}{a}\partial_a\biggl) + \frac{1}{2\omega + 3} a\phi\partial_a\partial_\phi
- \frac{1}{2\omega + 3} \phi^2\biggr(\partial^2_\phi + \frac{q}{\phi}\partial_\phi\biggl) -
\frac{1}{2}\phi^2\partial^2_\chi\biggl\}\Psi = 48ka^4\phi^2\Psi \quad ,
\end{equation}
where $p$ and $q$ are factor ordering terms.
\par
The equation (\ref{wdw1}) has mixed second order derivatives in the
variables $a$ and $\phi$; at the
same time, the potential in the right hand side mixes these same
variables. In this form, the equation
(\ref{wdw1}) can not be solved through the method of separation of
variables.
However, we can redefine $a$ and $\phi$, writing
\begin{equation}
\label{ct}
a = \phi^{-\frac{1}{2}}b \quad , \phi' = \phi \quad , \chi' = \chi.
\end{equation}
This transformation is the minisuperspace version of the transformation
that takes Lagrangian (\ref{l22}), expressed in the so-called Jordan's
frame, to a new Lagrangian
where gravity is coupled minimally to two scalar fields.
Inserting the transformation (\ref{ct}) in (\ref{wdw1}), we obtain a new
equation for the dynamical variables
$b$ and $\phi$:
\begin{equation}
\label{wdw2}
\biggr\{\frac{b^2}{12}\biggr(\partial^2_b + \frac{p'}{b}\partial_b\biggl)
-
\frac{\phi^2}{3 + 2\omega}\biggr(\partial^2_\phi +
\frac{q'}{\phi}\partial_\phi\biggl) - 
\frac{\phi^2}{2}\partial_\chi\biggl\}\Psi = 48b^4\Psi \quad .
\end{equation}
The new factor order terms $p'$ and $q'$ are related to the older ones,
$p$ and $q$, by the expressions,
\begin{equation}
p' = \frac{12}{3 + 2\omega}\biggr(\frac{\omega}{6}p + \frac{3}{4} -
\frac{q}{2}\biggl) \quad ,
\quad q' = q \quad .
\end{equation}
When $p = q = 1$ then $p' = q' = 1$. From this transformation, we can
deduce that the Wheeler-DeWitt equation
for a minimal coupled gravity/scalar field is equivalent to a non minimal
coupled version of this model.
The conformal equivalence at the classical level is preserved at the quantum
level.
Moreover, the factor ordering is preserved only when $p = q = 1$
because in this case the differential operator in the Wheeler-DeWitt equation is the Laplacian
one, which is covariant under general field redefinitions. 
\par
Equation (\ref{wdw2}) has already been studied in the
literature \cite{nelson1}. The solutions were found using the separation
of variables method. To obtain the solutions of Eq. (\ref{wdw1})
(which is in the Jordan's
frame), we have just to employ the transformation (\ref{ct}) in the solutions so found. Writing $\Psi(b,\phi,\chi) =\rho(b)\beta(\phi)\gamma(\chi)$, they read
\begin{eqnarray}
\label{eigen1}
\rho (b=a\phi ^{1/2}) &=& 
a^{(1-p')/2}\biggr[A_\rho I_n(12a^2\phi) + B_\rho
K_n(12a^2\phi)\biggl] \quad ,  \\\quad
n &=& \frac{\sqrt{(p' - 1)^2 - 48k_2}}{4} \quad ,\nonumber \\
\nonumber\\
\label{eigen2}
\beta (\phi) &=& \phi^{(1-q')/2}\biggr[A_\beta I_m(2\sqrt{(3 + 2\omega)k_1}\phi)
+ B_\beta K_m(2\sqrt{(3 + 2\omega)k_1}\phi)\biggl] \quad ,\\
 m &=& \frac{\sqrt{(q' - 1)^2 - 4(3 + 2\omega)k_2}}{2}\nonumber
 \quad , \\
\nonumber\\
\label{eigen3}
\gamma (\chi) &=& A_\gamma\exp(i\sqrt{8k_1}\chi) +
B_\gamma\exp(-i\sqrt{8k_1}\chi) \quad ,
\end{eqnarray}
where $k_1$ and $k_2$ are two constants of separation.
The general solution can be written as
\begin{equation}
\label{flu}
\Psi = \int
A(k_1,k_2)\rho_{k_1}(a,\phi)\beta_{k_2,k_1}(\phi)\gamma_{k_1}(\chi)dk_1dk_2
\quad .
\end{equation}

\section{The quantum bohmian trajectories}

In this section, we will apply the rules of the causal interpretation to
the
wave functions we have obtained in the previous section. We first summarize these
rules for the case of homogeneous minisuperspace models.
The general minisuperspace Wheeler-DeWitt equation is:
\begin{equation} 
\label{bsc}
{\cal H}({\hat{p}}^{\alpha}(t), {\hat{q}}_{\alpha}(t)) \Psi (q) = 0.
\end{equation}
where $p^{\alpha}(t)$ and $q_{\alpha}(t)$ represent the homogeneous 
degrees of freedom coming from the gravitational and matter degrees of freedom.
Writing $\Psi = R \exp (iS/\hbar)$, and substituting it into (\ref{bsc}),
we obtain the following equation:
\begin{equation}
\label{hoqg}
\frac{1}{2}f_{\alpha\beta}(q_{\mu})\frac{\partial S}{\partial q_{\alpha}}
\frac{\partial S}{\partial q_{\beta}}+ U(q_{\mu}) +
Q(q_{\mu}) = 0,
\end{equation}
where
\begin{equation}
\label{hqgqp}
Q(q_{\mu}) = -\frac{1}{R} f_{\alpha\beta}\frac{\partial ^2 R}
{\partial q_{\alpha} \partial q_{\beta}},
\end{equation}
and $f_{\alpha\beta}(q_{\mu})$ and $U(q_{\mu})$ are the minisuperspace
version of the DeWitt metric \cite{dew} and of the
scalar curvature density of the spacelike
hypersurfaces,
respectively.
The causal interpretation applied to quantum cosmology states that
the trajectories $q_{\alpha}(t)$ are real, independently of any
observations.
Eq. (\ref{hoqg}) is the Hamilton-Jacobi equation for them, which
is the classical one
amended with a quantum potential term (\ref{hqgqp}), responsible
for the quantum effects. This suggests
to define:
\begin{equation}
\label{h}
p^{\alpha} = \frac{\partial S}{\partial q_{\alpha}} ,
\end{equation}
where the momenta are related to the velocities in the usual way:
\begin{equation}
\label{h2}
p^{\alpha} = f^{\alpha\beta}\frac{1}{N}\frac{\partial q_{\beta}}{\partial
t} \; .
\end{equation}
To obtain the quantum trajectories we have to solve the following
system of first order differential equations:
\begin{equation}
\label{h3}
\frac{\partial S(q_{\alpha})}{\partial q_{\alpha}} =
f^{\alpha\beta}\frac{1}{N}\frac{\partial q_{\beta}}{\partial t} \; .
\end{equation}

Eqs. (\ref{h3}) are invariant under time reparametrization. Hence,
even at the quantum level, different choices of $N(t)$ yield the same 
spacetime geometry for a given non-classical solution $q_{\alpha}(t)$.
There is no problem of time in the causal interpretation of minisuperspace
quantum cosmology. The classical limit is given by the region where
$Q=0$ and the Hamilton-Jacobi equation (\ref{hoqg}) reduces to the classical
one.

Let us then apply this interpretation to our minisuperspace models
and
choose the gauge $N = 1$.
The bohmian
trajectories can be obtained by integrating
the relations,
\begin{eqnarray}
\pi_a = \partial_a S \quad ,\\
\pi_\phi = \partial_\phi S \quad , \\
\pi_\chi = \partial_\chi S \quad ,
\end{eqnarray}
which lead to the differential equations,
\begin{eqnarray}
\label{ba}
24a\dot a\phi + 12a^2\dot\phi &=& \partial_a S \quad , \\
\label{bf}
12a^2\dot a - 4\omega\frac{\dot\phi}{\phi}a^3 &=& \partial_\phi S \quad , \\
\label{bx}
- 4\frac{a^3}{\phi}\dot\chi &=& \partial_\chi S \quad .
\end{eqnarray}
Employing the conformal transformations (\ref{ct}) and remembering that $dt =
\phi^{-1/2}d\tau$, where $\tau$ is the proper time in Einstein's frame,
we obtain the
differential equations
\begin{eqnarray}
\label{pht1}
24bb' &=& \partial_b S \quad , \\
\label{pht2}
-2(3 + 2\omega)b^3\frac{\phi'}{\phi^2} &=& \partial_\phi S \quad , \\
\label{pht3}
- 4\frac{b^3}{\phi ^2}\chi' &=& \partial_\chi S \quad .
\end{eqnarray}
The primes here mean derivatives with respect to the proper time in the
Einstein frame.
Relations (\ref{pht1},\ref{pht2},\ref{pht3}) are the same that were found in the
case of a minimal coupling
between gravity and a scalar field.
\par
Inserting in the Wheeler-DeWitt equation the expression $\Psi = R \exp (iS/\hbar)$, we
find the corresponding
Hamilton-Jacobi equation, which reads for the non minimal coupling case,
\begin{equation}
\label{hj}
- \frac{\omega a^2}{6(2\omega + 3)}(\partial_aS)^2 - \frac{a\phi}{2\omega + 3} (\partial_aS)(\partial_\phi S)
+ \frac{\phi ^2}{2\omega + 3}(\partial_\phi S)^2
+ \frac{\phi ^2}{2}(\partial_\xi S)^2 + Q + V_{cl} = 0 \quad ,
\end{equation}
where
\begin{eqnarray}
\label{pc}
V_{cl} &=& - 48ka^4\phi^2 \quad , \\
\label{pq}
\nonumber
Q &=& \frac{1}{R}\biggr[\frac{\omega a^2}{6(2\omega + 3)}\biggr(\partial^2_aR +
\frac{p}{a}\partial_aR\biggl) + \frac{a\phi}{2\omega + 3}
\partial^2_{\phi a}R  \\ 
& & - \frac{\phi ^2}{2\omega + 3}
\biggr(\partial^2_\phi R + \frac{q}{\phi}\partial_\phi R\biggl) - 
\frac{\phi ^2}{2}\partial^2_\xi R\biggl] \quad , 
\end{eqnarray}
where $V_{cl}$ and $Q$ are the classical and quantum potentials. 
The transformations (\ref{ct}) map  equation (\ref{hj}) and the
potentials (\ref{pc},\ref{pq}) into
the corresponding Hamilton-Jacobi equation, classical and
quantum potentials of the minimal
coupling case \cite{nelson1}.
This complete the conformal equivalence (from the mathematical point
of view) between minimal and non-minimal coupling at the quantum level
in the causal interpretation.
\par
In \cite{nelson1}, we have obtained the bohmian trajectories by integrating Eqs.
(\ref{pht1},\ref{pht2},\ref{pht3}) for some exact solutions of the Wheeler-DeWitt
equation, and we have
found that the classical limit
is recovered when the scale factor is small. 
With these results, we can obtain the bohmian trajectories by using the
conformal transformations (\ref{ct}), and investigate the presence of
singularities and classical limit of these solutions
in the non minimal coupling case.
\par 
First, we remember the expressions connecting the scale factor and proper
time
in the minimal and non minimal coupling cases: $a = b\phi^{-1/2}$ and $t =
\int \phi^{-1/2}d\tau$.
The classical limit in the minimal case is
\begin{equation}
\label{min}
b \sim \tau^{\frac{1}{3}}, \quad \phi \sim \tau^{\frac{2}{3\alpha}}, \quad
\chi = {\rm const.}.
\end{equation}
Using the conformal transformation, we obtain the classical limit for the
non minimal coupling,
\begin{equation}
\label{non}
a \propto t^\frac{\alpha - 1}{3\alpha - 1}, 
\quad \phi \propto t^\frac{2}{3\alpha- 1},
\quad \chi = {\rm const.}.
\end{equation}
where $\alpha = \sqrt{1 + \frac{2}{3}\omega}$, which were obtained directly
from the classical solutions in Jordan's frame and shown in Eq. (\ref{roni}). 
In the Einstein's frame, the qualitative behavior of the scale factor is the same
irrespective of the value of $\omega$, while in the Jordan's frame its behavior
depends crucially on the value of $\omega$, as explained in the section (2).
In order to analyze all these possibilities at the quantum level,
we will study four examples. The solutions obtained in cases (I,II,III)
where extracted from Ref. \cite{gra}.
\vspace{0.8cm}
\newline
Case I: $\omega=9/2$ ($\omega > 0$), which yields $\alpha=2$. 

We will choose
$A(k_1,k_2) = \frac{3}{2}\delta(k_1 -
\exp (i\frac{\pi}{3})/48)\sinh(\pi\nu)$, 
with $\nu \equiv \sqrt{3k_2}$, $A_{\rho}=A_{\beta}=B_{\gamma}=0$, and
$p'=q'=1$ in Eq. (\ref{flu}).
\vspace{0.5cm}
\newline
1) The minimal coupling case.

In this case, the wavefunction takes the form,
\begin{equation}
\Psi = \frac{\pi ^{3/2}}{2^{5/2}}\frac{\phi}{\sqrt{y}}\exp\biggr(-y-\frac{\phi
^2}{16y}-\frac{\chi}{2\sqrt{6}}\biggl)
\exp\biggr[i\biggr(\frac{\pi}{6} - \frac{\sqrt{3}\phi ^2}{16y} +
\frac{\chi}{2\sqrt{2}}\biggl)\biggl] \quad .
\end{equation}
where $y \equiv 12b^2$.
We obtain the following expressions for the bohmian trajectories:
\begin{eqnarray}
b &\propto& \tau ^{1/3} \quad \mbox{or} \quad y \propto \tau ^{2/3}\quad , \\
\phi &\propto&  \tau ^{1/3},\\
\chi &\propto&  \tau ^{2/3} + \chi _0 \quad .
\end{eqnarray}
They coincide with the classical trajectories for $\tau,b,\phi$ very small
(see Eq. (\ref{min})) but they differ from them when these conditions are not satisfied
(see Eqs. (\ref{cs1},\ref{cs2},\ref{cs3})).
The quantum potential is 
\begin{equation}
Q = \frac{y^2}{3}-\frac{\phi ^2}{16} \quad ,
\end{equation}
while the classical potential is
\begin{equation}
V_{cl} = -\frac{y^2}{3} \quad .
\end{equation}
The kinetic terms are
\begin{eqnarray}
K_b &=& -\frac{b^2}{12}\biggr(\frac{\partial
S}{\partial
b}\biggl)^2 \propto \frac{\phi^4}{y^2} \quad , \\
K_{\phi} &=& 
\frac{\phi ^2}{2}\biggr(\frac{\partial
S}{\partial\phi}\biggl)^2 \propto \frac{\phi^4}{y^2} \quad , \\
K_{\chi} &=& 
\frac{\phi ^2}{12}\biggr(\frac{\partial
S}{\partial\chi}\biggl)^2 \propto \phi^2 \quad .
\end{eqnarray}
It can be seen from the above expressions that, for small $\tau$, 
$K_b$ and $K_{\phi}$ dominates
while, for large $\tau$, $Q$ and $V_{cl}$ dominates ($Q/V_{cl}\approx 1$). 
In other words, quantum effects are negligible for small $b$ and $\phi$
but become important otherwise.
\vspace{0.5cm}
\newline
2) The non minimal coupling case.

To obtain the new wavefunction we have just to make the substitution
$y=x\phi$ where $x=12a^2$, yielding
\begin{equation}
\Psi = \frac{\pi ^{3/2}}{2^{5/2}}\sqrt{\frac{\phi}{x}}\exp\biggr(-x\phi-\frac{\phi}{16x}-
\frac{\chi}{2\sqrt{6}})
\exp\biggr[i\biggr(\frac{\pi}{6} - \frac{\sqrt{3}\phi}{16x} +
\frac{\chi}{2\sqrt{2}}\biggl)\biggl] \quad .
\end{equation}
The bohmian trajectories can be calculated either by solving equations
(\ref{ba},\ref{bf},\ref{bx}) directly, or by making the conformal
mapping $x=y/\phi,\phi=\phi,\chi=\chi$ in the solutions of the minimal
coupling case.
We obtain the following expressions for the bohmian trajectories:
\begin{eqnarray}
a &\propto& t ^{1/5} \quad \mbox{or} \quad x \propto t ^{2/5}\quad , \\
\phi &\propto&  t ^{2/5},\\
\chi &\propto&  t ^{4/5} + \chi _0 \quad .
\end{eqnarray}
They coincide with the classical trajectories for $t,b,\phi$ very small
(see Eq. (\ref{non})) but they differ from them when these conditions are not satisfied.
The quantum potential is 
\begin{equation}
Q = \frac{x^2\phi ^2}{3}-\frac{\phi ^2}{16} \quad ,
\end{equation}
while the classical potential is
\begin{equation}
V_{cl} = -\frac{x^2\phi^2}{3} \quad .
\end{equation}
The kinetic terms are
\begin{eqnarray}
K_a &=& -\frac{a^2}{16}\biggr(\frac{\partial
S}{\partial
a}\biggl)^2 \propto \frac{\phi^2}{x^2} \quad , \\
K_{\phi} &=& 
\frac{\phi ^2}{12}\biggr(\frac{\partial
S}{\partial\phi}\biggl)^2 \propto \frac{\phi^2}{x} \quad , \\
K_{a\phi} &=& -
\frac{a\phi ^2}{12}\frac{\partial S}{\partial a}\frac{\partial
S}{\partial\phi}\propto \frac{\phi^2}{x^2} \quad , \\
K_\chi &=& 
\frac{\phi ^2}{2}\biggr(\frac{\partial
S}{\partial\chi}\biggl)^2 \propto \phi^2 \quad .
\end{eqnarray}
It can be seen from the above expressions that, for small $t$, 
$K_a$ and $K_{a\phi}$ dominates
while, for large $t$, $Q$ and $V_{cl}$ dominates ($Q/V_{cl}\approx 1$). 
In other words, quantum effects are negligible for small $a$ and $\phi$
but become important otherwise. 
\vspace{0.8cm}
\newline
Case II: $w=0$ which yields $\alpha=1$. 

We will choose 
$A(k_1,k_2) = \frac{3}{2}\delta(k_1 +
\frac{1}{12})\tanh(\pi\nu)$, with $\nu \equiv \sqrt{3k_2}$, $A_{\rho}=A_{\beta}=0, A_{\gamma}=B_{\gamma}=1$, and
$p'=q'=1$ in Eq. (\ref{flu}).
\vspace{0.5cm}
\newline
1) The minimal coupling case.

In this case, the wavefunction takes the form,
\begin{equation}
\label{flup}
\Psi = \frac{\pi}{2}\cosh(\sqrt{\frac{2}{3}}\chi)\sqrt{\frac{y\phi}{y^2 +
\phi^2}}e^{-y}
\exp\biggr[i\biggr(\frac{\pi}{4} - \phi -
\arctan\biggr(\frac{\phi}{y}\biggl)\biggl)\biggl] \quad .
\end{equation}
The bohmian trajectories are (see Ref. \cite{nelson1}):
\begin{equation}
\label{s3}
\begin{array}{l}
b = \frac{1}{\sqrt{12}}\biggr[\ln\biggr(\frac{C}{\sqrt{1 +
4\eta ^2}}\biggl)\biggr]^{\frac{1}{2}} \quad , \\
\phi = - \frac{1}{2\eta}\ln\biggr(\frac{C}{\sqrt{1 + 4\eta ^2}}\biggl) =
- 6\frac{a^2}{\eta} \quad ,\\
\chi = {\rm const.} \quad ,
\end{array}
\end{equation}
where $\eta = \int\frac{d\tau}{b}$ is the conformal time and $C$ is an
integration
constant. 
For small $b$, when $\eta$ approaches $\pm \frac{\sqrt{c^2 -1}}{2}$, these
functions
tend to:
\begin{equation}
\begin{array}{l}
b(\tau) \propto \tau ^{\frac{1}{3}} \quad , \\
\phi(\tau) \propto \tau ^{\frac{2}{3}} \propto b^2 \quad ,\\
\chi = {\rm const.} \quad .
\end{array}
\end{equation}
which is exactly the classical behavior for $\omega = 0$. When $b$ is not
small, the trajectories are not classical. This can be explained by the
behavior of the quantum potential
\begin{equation}
Q = \frac{1}{3}\frac{(y^4 - 2\phi^2 y - \phi^4)}{y^2 + \phi^2} \quad .
\end{equation}
When $b$ is small the kinetic terms dominate while for $b$ large
the quantum and classical potential dominate (see Ref. \cite{nelson1}
for details).
\vspace{0.5cm}
\newline
2) The non minimal coupling case.

A conformal transformation in the solution (\ref{flup}) yields
\begin{equation}
\Psi = \frac{\pi}{2}\cosh(\frac{2}{3}\chi)\sqrt{\frac{x}{x^2 +
1}}e^{-x\phi}
\exp\biggr[i\biggr(\frac{\pi}{4} - \phi -
\arctan\biggr(\frac{1}{x}\biggl)\biggl)\biggl] \quad .
\end{equation}
The bohmian trajectories can be calculated either by solving equations
(\ref{ba},\ref{bf},\ref{bx}) directly or by making the conformal
mapping $x=y/\phi,\phi=\phi,\chi=\chi$ in the solutions of the minimal
coupling case.
The new solutions are
\begin{equation}
\label{s4}
\begin{array}{l}
a = \biggr(-\frac{t}{4}\biggl) ^{1/3} \quad , \\
\phi = - \frac{1}{3(-2t)^{2/3}}\ln\biggr(\frac{C}{\sqrt{1 + 
9(-2t) ^{4/3}}}\biggl) \quad ,\\
\xi = {\rm const.} \quad ,
\end{array}
\end{equation}
They have an initial singularity, and they are completely different from the classical trajectories, which
in this case are oscillatory and without singularities.
This is an example where quantum effects can create a singularity. 
There is no classical limit. This can be seen by examining the behavior of
the quantum potential
\begin{equation}
Q = \frac{1}{3}\frac{(\phi ^2 x^4 - 2\phi^2 x - \phi^2)}{x^2 + 1} \quad .
\end{equation}
when compared with the classical potential and kinetic terms
\begin{eqnarray}
V_{cl} &=& -\frac{x^2\phi^2}{3} \quad , \\
K_a &=& 0\quad , \\
K_{\phi} &=& 
\frac{\phi ^2}{3}\biggr(\frac{\partial
S}{\partial\phi}\biggl)^2 \propto \phi^2 \quad , \\
K_{a\phi} &=& -
\frac{a\phi ^2}{3}\frac{\partial S}{\partial a}\frac{\partial
S}{\partial\phi}\propto \frac{x\phi}{1+x^2} \quad , \\
K_\chi &=& 
\frac{\phi ^2}{2}\biggr(\frac{\partial
S}{\partial\chi}\biggl)^2 = 0 \quad .
\end{eqnarray}
When $t$ is small, $Q$ and $K_{\phi}$ dominate while for $t$ large
$Q$ and $V_{cl}$ dominate.
\vspace{0.8cm}
\newline
Case III: $\omega=-9/8$ ($-4/3 < \omega < 0$), which yields $\alpha=1/2$. 

We will choose
$A(k_1,k_2) = \frac{3}{2}\delta(k_1 - \frac{1}{3}
\exp (i\frac{\pi}{3}))\sin(\frac{\pi\nu}{2})$, 
with $\nu \equiv \sqrt{3k_2}$, $A_{\rho}=A_{\beta}=B_{\gamma}=0$, and
$p'=q'=1$ in Eq. (\ref{flu}).
\vspace{0.5cm}
\newline
1) The minimal coupling case.

In this case, the wavefunction takes the form,
\begin{equation}
\Psi = \frac{\pi ^{3/2}}{2^{1/2}}\frac{y}{\sqrt{\phi}}\exp\biggr[-\frac{\sqrt{3}}{2}\biggr(\phi+
\frac{y^2}{8\phi}\biggl)-\frac{2\chi}{\sqrt{6}}\biggl]
\exp\biggr[i\biggr(-\frac{\pi}{12} - \frac{\phi}{2} + \frac{y^2}{16\phi} +
2\chi\biggl)\biggl] \quad .
\end{equation}
We obtain the following expressions for the bohmian trajectories:
\begin{eqnarray}
b &=& \lambda _1\sqrt{\mbox{sech}(\lambda _1^2\epsilon)}
\quad , \\
\phi &=& 6\lambda _1^2\mbox{cosech}(2\lambda _1^2\epsilon) \quad ,
\\
\chi &=& 9\lambda _1^2\mbox{cotanh}(2\lambda _1^2\epsilon) + \chi _0  \quad ,
\end{eqnarray}
where $d\tau = b^3 d\epsilon$.
They coincide with the classical trajectories for $\epsilon$ very large
and $b,\phi$ very small, where they present the behavior of
Eq. (\ref{min}), but they differ from them when these conditions are not satisfied.
The quantum potential is 
\begin{equation}
Q = \frac{y^2}{4}-\frac{4\phi ^2}{3} \quad ,
\end{equation}
while the classical potential, is as usual,
\begin{equation}
V_{cl} = -\frac{y^2}{3} \quad .
\end{equation}
The kinetic terms are
\begin{eqnarray}
K_b &=& -\frac{b^2}{12}\biggr(\frac{\partial
S}{\partial
b}\biggl)^2 \propto \frac{y^4}{\phi^2} \quad , \\
K_{\phi} &=& 
\frac{4\phi ^2}{3}\biggr(\frac{\partial
S}{\partial\phi}\biggl)^2 = - \frac{1}{3} \biggr(\phi ^2 + 
\frac{y^2}{4} - \frac{y^4}{64 \phi ^2}\biggl) \quad , \\
K_\chi &=& 
\frac{\phi ^2}{2}\biggr(\frac{\partial
S}{\partial\chi}\biggl)^2 \propto \phi^2 \quad .
\end{eqnarray}
It can be seen from the above expressions that, for large $\epsilon$, 
$K_b$ and $K_{\phi}$ dominates
while, for $\epsilon << 1$, $Q$, $K_{\phi}$ and $K_{\chi}$ dominates. 
Quantum effects are negligible for small $b$ and $\phi$
but become important otherwise.
\vspace{0.5cm}
\newline
2) The non minimal coupling case.

To obtain the new wavefunction we have just to make the substitution
$y=x\phi$ where $x=12a^2$, yielding
\begin{equation}
\Psi = \frac{\pi ^{3/2}}{2^{1/2}}
x\sqrt{\phi}\exp\biggr[-\frac{\sqrt{3}}{2}\biggr(\phi+
\frac{x^2\phi}{8}\biggl)-\frac{2\chi}{\sqrt{6}}\biggl]
\exp\biggr[i\biggr(-\frac{\pi}{12} - \frac{\phi}{2} + \frac{x^2\phi}{16} +
2\chi\biggl)\biggl] \quad .
\end{equation}
We obtain the following expressions for the bohmian trajectories:
\begin{eqnarray}
a &=& \sqrt{\frac{\mbox{sinh}(\lambda _1^2\epsilon)}{3}}
\quad , \\
\phi &=& 6\lambda _1^2\mbox{cosech}(2\lambda _1^2\epsilon) \quad ,
\\
\chi &=& 9\lambda _1^2\mbox{cotanh}(2\lambda _1^2\epsilon) + \chi _0  \quad ,
\end{eqnarray}
where d$t=\phi a^3$d$\epsilon$ and we will define $t_{\infty}\equiv t(\epsilon
= \infty)$.
They coincide with the classical trajectories for $\epsilon$,$a$
very large, and $\phi$ very small. Their behaviors in the limit
$t\rightarrow t_{\infty}$ are
\begin{eqnarray}
a &\propto& (t_{\infty}-t) ^{-1} \quad \mbox{or} \quad
x \propto (t_{\infty}-t) ^{-2}\quad , \\
\phi &\propto&  (t_{\infty}-t) ^4,\\
\chi &\propto& \chi _0 \quad .
\end{eqnarray}
which coincide with Eq. (\ref{non}) for $\alpha =1/2$, but they differ from them when these conditions are not satisfied. For instance, when $\epsilon$,
and $t$, are very small they are
\begin{eqnarray}
a &\propto& t^{1/3} \quad \mbox{or} \quad 
x \propto t^{2/3}\quad , \\
\phi &\propto&  t^{-2/3},\\
\chi &\propto& t^{-2/3} \quad .
\end{eqnarray}
which is not the classical behavior in this case. Note that one
of the singularities of the classical solution is changed to a big-bang
or a big-crunch.
The quantum potential is 
\begin{equation}
Q = \frac{y^2\phi^2}{4}-\frac{4\phi ^2}{3} \quad ,
\end{equation}
while the classical potential is 
\begin{equation}
V_{cl} = -\frac{y^2}{3} \quad .
\end{equation}
The kinetic terms are
\begin{eqnarray}
K_a &=& -\frac{a^2}{4}\biggr(\frac{\partial
S}{\partial
a}\biggl)^2 \propto \phi x^3 \quad , \\
K_{\phi} &=& 
\frac{4\phi ^2}{3}\biggr(\frac{\partial
S}{\partial\phi}\biggl)^2 \propto 
\frac{\phi^2}{2}\biggr(1-\frac{x^2}{8}\biggl)  \quad , \\
K_{a\phi} &=& -
\frac{a\phi ^2}{9}\frac{\partial S}{\partial a}\frac{\partial
S}{\partial\phi}\propto \frac{\phi^2 x^2}{8}\biggr(1-\frac{x^2}{8}\biggl) 
\quad , \\
K_\chi &=& 
\frac{\phi ^2}{2}\biggr(\frac{\partial
S}{\partial\chi}\biggl)^2 \propto \phi^2 \quad .
\end{eqnarray}
It can be seen from the above expressions that, for $t\rightarrow t_{\infty}$, 
$K_a$ dominates
while, for small $t$, $Q$, $K_{\phi}$ and $K_{\chi}$ dominate. 
The quantum effects are negligible for large $a$ and small $\phi$
but become important otherwise. 
\vspace{0.8cm}
\newline
Case IV: $-3/2 < \omega \leq -4/3$, which yields $\alpha\leq 1/3$.

In this case, we do not take superpositions, but the wave functions
(\ref{eigen1},\ref{eigen2},\ref{eigen3}) themselves. We will choose
$B_{\rho}=B_{\beta}=B_{\gamma}=0$, and
$p'=q'=1$ yielding
\begin{equation}
\Psi = I_n(x\phi) I_m(2\sqrt{(3 + 2\omega)k_1}\phi) \exp (i\sqrt{8k_1}\chi)
\end{equation}
with $n=i\nu$ and $m = i\sqrt{(3 + 2\omega)k_2}$.
For small $\phi$ we can approximate the Bessel functions for
\begin{equation}
\Psi = C_0 \exp [i(\nu \ln (x\phi) + 
\mu \ln (2\sqrt{(3 + 2\omega)k_1}\phi) + \sqrt{8k_1}\chi)]
\end{equation}
As $C_0$ is a constant, the quantum potential is null and
the bohmian trajectories are the classical ones in this limit.
One can verify this by inserting the phase of the above wave function
into Eqs.
(\ref{ba},\ref{bf},\ref{bx}), obtaining 
\begin{equation}
a \propto t^\frac{\alpha - 1}{3\alpha - 1} \quad , \quad \phi \propto t^\frac{2}{3\alpha
- 1} \quad ,
\end{equation}
with $\alpha = \sqrt{1 + \frac{2}{3}\omega}$ as before.
Note that even in the cases where $a$ becomes large, the product
$x\phi \propto t^{\frac{2\alpha }{3\alpha - 1}}$ is always small,
justifying our approximation. We can see that the classical
limit happens for $\phi$ small, but it can happen for large $a$.

\section{Conclusions}
The possible avoidance of classical singularities due to quantum
effects, and the prediction of the classical behaviour of the 
Universe are some of the most
important issues in quantum cosmology.
These questions have been studied by different methods in the literature. In
Ref. \cite{nelson1}, it has been proposed
to use the bohmian trajectories to study these issues, and it was verified
that, in the context of scalar-tensor theories coming from string and
Kaluza-Klein theories, re-expressed in the
Einstein's frame, their conclusions agrees with usual semi-classical
considerations. The results obtained for some exact solutions of the
Wheeler-DeWitt equation indicated that the Universe is
classical when the scale factor is small
(comparable to Planck scale), and the singularities are not avoided. In the present work we have extended this analysis to non minimal coupled
scalar-tensor theories. The conformal
equivalence between Jordan's and Einstein's frame at classical and
quantum level was established, even for the bohmian trajectories and quantum potentials.
Having the particular exact solutions of the Wheeler-DeWitt equation
in the Einstein's frame, we were able to obtain
the corresponding solutions in the Jordan's frame.
The bohmian trajectories in Jordan's frame present classical
behavior whenever the scalar field non minimally coupled with gravity
is small, which coincides with small 
values of the scale factor $a$ when $\omega > 0$, and large values of $a$ when 
$-\frac{3}{2} < \omega  < 0$). Quantum behavior appears when
the scalar field is not small, which means for large values of 
$a$ when $\omega > 0$, and for small values of $a$ when 
$-\frac{3}{2} < \omega  < 0$). Hence, contrary to the minimal
coupling case,
the bohmian trajectories may coincide with the classical solutions for
large values of $a$.
The exception is the very particular case $\omega =0$. In this case, the classical solution in Jordan's frame is non-singular and periodic. 
In an example studied
in this paper, the corresponding bohmian trajectory has always quantum
behavior, even when the scalar field becomes small. More interesting,
this quantum solution presents an initial singularity showing that,
depending on the quantum state of the system, quantum effects, rather than avoid, may create singularities where classically there is none.

These results show that the quantum properties of quantum minisuperspace
solutions may be quite different in Einstein's and Jordan's frame, although
they may be linked by a conformal transformation. Those frames have not the
same physical content, and the equivalence is manifest only from the
mathematical point of view.
A simple example
is the fact that a G-variable theory demands the Jordan's frame, while a
G-constant is a natural feature
of the Einstein's frame. Also, our results express the importance of 
the choice of boundary conditions for the Wheeler-DeWitt equation. Many
solutions we have obtained here present unphysical features, like quantum
behavior for large scale factors. It should be interesting to investigate
other solutions of the Wheeler-DeWitt equation presented in this paper which
may be not expressible in term of elementary functions, but which present
more appealing features, like absence of singularities, classical limit
for large $a$, and inflation. One possibility is to study gaussian 
superpositions of the Bessel functions presented in section 3, and study
them numerically. As gaussian superpositions mixes negative and positive 
indices of the Bessel functions (here we made superpositions with only one
sign of the indices because these are the ones which can be expressed in
term of elementary functions), and as these indices are connected with expansion
and contraction, it may be possible to obtain non-singular quantum solutions
in both frames. It should also be interesting to extend the present analysis
to any value of the the curvature of the spatial sections. These are
subjects of our future investigations. 
\vspace{1.0cm}

{\bf ACKNOWLEDGEMENTS}

We would like to thank the Cosmology Group of CBPF for useful discussions.
We would also like to thank CNPq of Brazil and the Brazilian-French
cooperation CAPES / COFECUB (project 180/96) for financial support.

\end{document}